# Magnetic 2D materials and heterostructures

M. Gibertini[1,2], M. Koperski[3,4], A. F. Morpurgo[1,5], K. S. Novoselov[3,4]

[1]*Department of Quantum Matter Physics, University of Geneva, CH-1211 Geneva, Switzerland*

[2]*National Centre for Computational Design and Discovery of Novel Materials (MARVEL), École Polytechnique Fédérale de Lausanne, CH-1015 Lausanne, Switzerland*

[3]*School of Physics and Astronomy, University of Manchester, Oxford Road, Manchester, M13 9PL, UK*

[4]*National Graphene Institute, University of Manchester, Oxford Road, Manchester, M13 9PL, UK*

[5]*Group of Applied Physics, University of Geneva, CH-1211 Geneva, Switzerland*

**The family of 2D materials grows day by day, drastically expanding the scope of possible phenomena to be explored in two dimensions, as well as the possible van der Waals heterostructures that one can create. Such 2D materials currently cover a vast range of properties. Until recently, this family has been missing one crucial member – 2D magnets. The situation has changed over the last two years with the introduction of a variety of atomically-thin magnetic crystals. Here we will discuss the difference between magnetic states in 2D materials and in bulk crystals and present an overview of the 2D magnets that have been explored recently. We will focus, in particular, on the case of the two most studied systems – semiconducting $CrI_3$ and metallic $Fe_3GeTe_2$ – and illustrate the physical phenomena that have been observed. Special attention will be given to the range of novel van der Waals heterostructures that became possible with the appearance of 2D magnets, offering new perspectives in this rapidly expanding field.**

# Introduction

The family of two-dimensional (2D) crystals has been growing with an impressive speed over the last few years, with many new systems have already been introduced[1-5], while many more have been predicted and await discovery[6-9]. The range of properties covered by such materials is now extremely large: metals, semimetals, topological insulators, semiconductors and insulators – all are present within this class of crystals[2,10]. The scope of correlation phenomena hosted in these systems is also rather broad, and includes superconductivity, charge density waves, Mott insulators, *etc*. More often than not, one atom thick crystals possess properties that are strikingly different from those of their three dimensional (3D) parent compounds: graphene is a zero gap semiconductor[11], whereas graphite is a semimetal with a band overlap; many monolayers of transition metal dichalcogenides (TMDC) in 2H phase are direct band gap semiconductors[12], whereas in their bulk form (and even in their bilayers) the band gap is indirect. Combinations of such materials with different properties have already led to the creation of unique heterostructures, allowed investigation of new physical effects, as well as the creation of novel devices[2,13,14].

A conspicuously missing member of the family of 2D materials are magnetic atomically-thin crystals[15]. Until recently 2D magnetism remained an elusive dream and such crystals were not available for experimental work. At the same time, the opportunities which would be opened by 2D magnets are huge, ranging from the possibility of exploring – or even exploiting – a plethora of different 2D magnetic states to that of controlling the magnetic properties, e.g. with an external electric field. Furthermore, using such materials as parts of van der Waals heterostructures is likely to disclose new exciting directions. For instance, stacking different magnetic crystals together –or restacking the same 2D crystals with different orientation– might result in a different magnetic order or in new physical phenomena.

Somehow, the absence of magnetic 2D materials was surprising, because layered, van-der-Waals-bonded, magnetic crystals have been available to us for a long time[16,17]. It should be possible to exfoliate them down to one atomically thin unit and potentially grow them on substrate as individual monolayers. Indeed, some theoretical efforts had been done to demonstrate that such crystals would be stable and exhibit a finite critical temperature[18-23] $T_c$.

It was only recently, however, that a number of 2D magnets have been obtained experimentally[24]. This breakthrough was immediately followed by intense experimental and theoretical activities to start exploring the field of 2D magnetism. The full arsenal of techniques developed for van der Waals heterostructures was applied to the study of such crystals and a number of specific effects have been observed: from renormalization of $T_c$ to the control of magnetic ordering by gating. Here we discuss which particular physical phenomena can manifest themselves differently in 2D (as compared to 3D) and summarize the current experimental situation in the field.

# From bulk to 2D magnetism

## Theoretical consideration

The hallmark of magnetism is the existence of an ordered arrangement of magnetic moments over macroscopic length scales, with a spontaneous breaking of time-reversal symmetry. This is typically driven by the interaction between the neighbouring spins (exchange coupling) that tends to favour specific relative orientations between them. At zero temperature ($T$), this local order can extend over macroscopic length scales. With increasing $T$, thermal fluctuations tend to misalign magnetic moments in neighbouring regions, so that long-range order is destroyed above $T_c$. Indeed, whether or not a system undergoes a phase transition at a finite $T_c$ depends on the effectiveness of thermal fluctuations, which is governed by few general parameters irrespective of specific details of the system. In particular, dimensionality plays an essential role in determining the impact of thermal fluctuations on the critical behaviour of many-body systems.

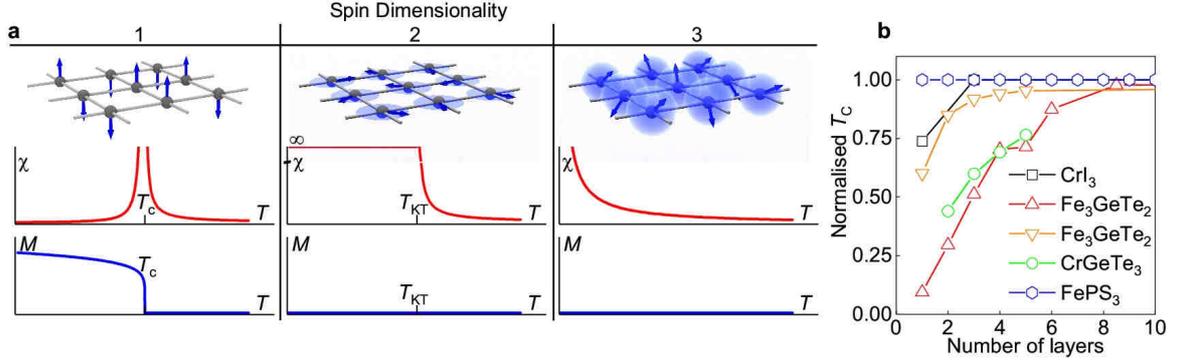

**Figure 1**. **Role of spin dimensionality and evolution of $T_c$. a**, A spin dimensionality $n=1$ means that the system has a strong uniaxial anisotropy and the spins point in either one of the two possible orientations ("up" or "down") along a given direction. The system behaves effectively as if it had only a single spin component along the easy axis and the underlying spin Hamiltonian for localized spins is called Ising model. The case $n=2$ corresponds to an easy-plane anisotropy that favours the spins to lie in a given plane, although the orientation within the plane is completely unconstrained. The spins can be thus considered to have effectively only two components (associated with the two in-plane directions), being successfully described within the so-called XY model. Note, that in this case $\chi \to \infty$ for $T<T_{KT}$. Finally, for isotropic systems $n=3$ and there is no constraint on the direction of the spins. The underlying spin Hamiltonian in this case is the isotropic Heisenberg model. **b**, $T_C$ (normalised to bulk $T_c^{3D}$ for the particular material) as a function of the number of layers. Data for $CrI_3$ are adapted from[47], for $Fe_3GeTe_2$ (red curve) – from[59], for $Fe_3GeTe_2$ (orange curve) – from[55], for $CrGeTe_3$ – from[48], for $FePS_3$ – from[42]

In a 3D system, a magnetic phase transition can always occur at a finite temperature, while in the one-dimensional case long-range order is possible only at $T = 0$[25]. Being at the border between these two extremes, the situation in 2D is more complex. In this case, the existence of magnetic long-range order at any finite temperature crucially depends on the number $n$ of relevant spin components, usually called spin dimensionality (Fig. 1), and determined by the physical parameters of the system (e.g., the presence and strength of magnetic anisotropy, see below).

Focusing for definiteness on finite-range exchange interactions, the Mermin-Wagner-Hohenberg theorem[26,27] states that thermal fluctuations destroy long-range magnetic order in 2D systems at any finite temperature when the spin dimensionality is three (isotropic Heisenberg model[28]). This is due to the fact that the continuous symmetry of the isotropic model leads to gapless long-wavelength excitations (spin waves) that have a finite density of states in 2D and can thus be easily excited at any finite temperature, with detrimental effects on magnetic order. On the contrary, the exact solution by Onsager[29] of the 2D Ising(-Lenz) model[30,31] shows that a phase transition to a magnetically ordered phase occurs at $T_c > 0$ when $n = 1$. In this case, the anisotropy of the system, which favours a specific spin component, opens a gap in the spin-wave spectrum thus suppressing the effect of thermal fluctuations.

For planar 2D magnets ($n=2$), conveniently described by the so-called XY model, there is no conventional transition to long-range order, although the susceptibility diverges below a finite temperature. Berezinskii[32], Kosterlitz and Thouless[33] pointed out that this divergence is associated with the onset of topological order, characterized by an algebraic decay of spin correlations and by the presence of bound pairs of vortex and antivortex arrangements of spins. Thus, below the so-called Kosterlitz-Thouless temperature $T_{KT}$, quasi-long-range magnetic order is established and the existence of a finite order parameter is suppressed only marginally with the system size.

In the absence of truly 2D magnetic crystals, such gamut of critical phenomena has been experimentally validated over the last thirty years in magnetic thin films grown on a substrate[34,35] or in 3D layered transition metal compounds[16,17], which can be viewed as a stack of weakly-coupled 2D magnetic layers. From these studies it clearly emerges that the theoretical scenarios presented above in terms of different spin dimensionalities can be considered as idealized limiting cases of a more realistic situation where spins have three components, but magnetic interactions restrict them to point mainly either along a given direction or in a specific plane.

A simple and yet successful strategy that allows interpolating between the above scenarios and to describe more realistic situations relies on the use of a generalized Heisenberg spin Hamiltonian:

$$H = -\frac{1}{2}\sum_{\langle i,j \rangle}\left(J\, \boldsymbol{S}_i \cdot \boldsymbol{S}_j + \Lambda\, S_i^z S_j^z\right) - \sum_i A\, (S_i^z)^2 \qquad (1)$$

Here $J$ is the exchange coupling between spins $\boldsymbol{S}_i$ and $\boldsymbol{S}_j$ on neighbouring sites (favouring either ferromagnetic, $J>0$, or antiferromagnetic, $J<0$, order), while $A$ and $\Lambda$ are the so-called on-site and inter-site (or exchange) magnetic anisotropies, respectively. This model captures the essential features of most experimental systems[17], and recovers in specific limits the idealized theoretical scenarios illustrated above. In particular, the isotropic Heisenberg model ($n$=3) corresponds to the absence of magnetic anisotropy ($A \approx 0$ and $\Lambda \approx 0$), while the Ising ($n$=1) or XY ($n$=2) model can be recovered in the limit of strong easy-axis or easy-plane anisotropy (e.g. $A \to +/-\infty$, respectively). This model can effectively take into account some subtle effects, as in the case of dipole-dipole interactions that can be partially described in terms of a renormalization of the on-site anisotropy $A$. Still, in some cases more sophisticated models are needed, which include either further neighbours or different kinds of magnetic coupling, such as Dzyaloshinsky–Moriya[36,37], Kitaev[38,39], or higher order (e.g. bi-quadratic) interactions.

A wide range of physical phenomena are expected depending on the values of $J, \Lambda,$ and $A$ or on the presence of additional magnetic interactions. The knowledge collected in the 1990s on magnetic layered van-der-Waals-bonded compounds, combined with the recent development on the production of monolayers by exfoliation or different growth methods, is now disclosing new opportunities. First, the rich variety of magnetic 2D materials that can be isolated offers the possibility to study model spin Hamiltonians in a broad range of parameter regimes, including regimes that have been remained so far unexplored. In addition, the generalised Heisenberg Hamiltonian (1) stems from a course-graining of the electronic properties, obtained by integrating out all degrees of freedom but spin. A change in the electronic structure, therefore, is expected to result in a modification of the effective spin Hamiltonian describing the system. This suggests that the substantial versatility of 2D materials and their sensitivity to external manipulations – gating, strain, coupling to other 2D materials in van der Waals heterostructures– provide unprecedented opportunities to further expand or tune the range of model parameters.

Whereas the approach based on the effective Hamiltonian in Eq. (1) is well suited to investigate the equilibrium configuration of 2D magnetic materials, it is far less ideal to understand the interplay of the magnetic states with the opto-electronic properties of atomically thin crystals, i.e. one of the domains in which magnetic 2D materials are anticipated to show new interesting physical phenomena and reveal unexpected results. The point is that the "effective" nature of the interaction hides the microscopic origin of the magnetic interaction parameters. Magnetic exchange integrals of the same magnitude could originate from direct exchange between nearby spins or from indirect mechanisms, such as super- or double- exchange, that are typically mediated by intermediate states involving non-magnetic atoms. Whereas the difference may be immaterial to determine the magnetic state in a system of given thickness, it is crucial to understand the evolution of the magnetic interactions upon reducing thickness. Direct exchange of two orbitals that are rather strongly localized in real space is unlikely to depend strongly on thickness. Conversely, indirect exchange mechanisms, being sensitive to the energy difference with the intermediate states and to the corresponding hopping energy, can be drastically altered as a result of the change in band structure that commonly happens in 2D materials when the thickness is reduced. Proper capturing of the nature of the coupling between the magnetic configuration and the single-electron wavefunctions is also clearly essential for our understanding of the influence of magnetism on such processes as tunnelling, inter-band radiative transition, and electron transport (and their evolution upon varying thickness). Early experiments (see below) are indeed showing that these processes are of great interest in 2D magnetic semiconductors, but very little is currently understood theoretically. There is plenty of room for theoretical work aiming at establishing the key concepts needed to properly capture the effect of the magnetic state on the semiconducting properties of magnetic 2D materials.

First efforts towards the isolation of magnetic 2D materials were carried out late in 2016 with the exfoliation of mono- and few-layers of $NiPS_3$ [40], $FePS_3$ [41-43], and $CrSiTe_3$ [44]. These compounds display magnetic order in bulk form[45,46], and Raman measurements provide –albeit only indirectly– evidence that magnetism is preserved in thin crystals. The major breakthrough occurred in 2017, with the first clear experimental confirmation of magnetism in atomically-thin $CrI_3$ [47] and $CrGeTe_3$ [48] down to the mono- and bi-layer limit, respectively.

These ground-breaking experiments have now sparked an intense activity on atomically-thin magnetic crystals[49]. Apart from rare exceptions[50-52], most efforts are focusing on the exfoliation of layered compounds that are known to be magnetic in their bulk form. Nonetheless, from the discussion above it should be clear that the inheritance of magnetism from 3D in the bulk to 2D in individual monolayers is not guaranteed a priori, and crucially depends on the magnetic anisotropy of the system. For instance, materials with isotropic exchange interactions can be magnetic in their 3D form, but thermal fluctuations strongly suppress the critical temperature as we reduce their thickness, until magnetism is completely destroyed in the 2D limit. This is the case of $CrGeTe_3$, for which the critical temperature goes to zero as we approach the monolayer limit[48]. The application of even a very small external magnetic field $B$ introduces an anisotropy in the system, opening up a gap in the spin-wave spectrum and thus allowing to have a non-zero magnetization at finite temperature. For $CrGeTe_3$, the authors of Ref. [48] introduce an effective critical temperature that in bilayers is already half the bulk $T_c$ (61 K) even just at $B$ = 0.065 T. On the contrary, when a material has already an intrinsic anisotropy, stemming, for instance, from a strong spin-orbit coupling[53,54] (magneto-crystalline anisotropy), magnetism is supposed to survive also in the monolayer limit (with the dimensional crossover from 3D to 2D leading to a different critical temperature, Fig. 1b). This is the case of $CrI_3$ and $Fe_3GeTe_2$, which will be discussed in details below. In addition to the spin-orbit coupling, other effects (e.g. magnetostatic dipole-dipole interactions) can contribute to a small magnetic anisotropy, yet sufficient to stabilize long-range magnetic order in 2D.

| Model | β $M(T<T_c) \propto |t|^\beta$ | γ $\chi \propto |t|^{-\gamma}$ | ν $\xi \propto |t|^{-\nu}$ | δ $M(T_c) \propto |B|^{1/\delta}$ | $T_c$ |
|---|---|---|---|---|---|
| 2D Ising | 1/8 | 7/4 | 1 | 15 | $2J/[k_B \ln(1+\sqrt{2})]$ Ref [29] |
| 3D Ising | 0.3265 | 1.237 | 0.630 | 4.789 | $T_c^{2D}(1+C(J_L/J)^{4/7})$ Ref [109] |
| 3D XY | 0.348 | 1.318 | 0.672 | 4.787 | $T_{KT} + C|J|/(\ln(|J/J_L|))^2$ Ref [110] |
| 3D Heisenberg | 0.369 | 1.396 | 0.711 | 4.783 | $C|J|/\ln(|J/J_L|)$ Ref [111,112] |
| Mean field | 0.5 | 1 | 0.5 | 3 | $zJ(1+z_L/z J_L/J)/k_B$ Ref [113] |

**Table 1. Critical behaviour from 2D to 3D.** Critical exponents governing the behaviour of magnetisation (*M*), susceptibility (*χ*), and correlation length (*ξ*) as a function of the reduced temperature, $t = T/T_c - 1$, or the magnetic field *B*, for different models in 2D and 3D. Not all the critical exponents are independent, as they are related by scaling relationships[114] such as $\gamma = \beta(\delta-1)$ and $2\beta + \gamma = d\nu$, where *d* is the space dimensionality. While the critical exponents for the 2D Ising model and mean-field theory can be determined analytically, for the other models they are results of numerical calculations[115]. In 2D, the critical exponents can be rigorously defined only for the Ising model, which is the only one displaying a standard phase transition to long-range order at finite $T_c$. Since long-range order in the 2D XY model is suppressed only in the thermodynamic limit, finite-size samples can sustain a non-vanishing magnetisation and its critical behaviour has been experimentally observed (e.g. in $Rb_2CrCl_4$ [116]), with a corresponding critical exponent[117] $\beta \approx 0.23$. For each model an expression for the critical temperature $T_c$ is also provided (*C* is a numerical coefficient and *z* and $z_L$ are the intra- and inter-layer coordination numbers, respectively), assuming in the 3D case a layered structure with $|J_L/J| \ll 1$ (see Box 1). In the case of the 2D Ising model, the argument of the logarithm depends on the lattice type, and the reported expression refers to a square lattice.

Given the crucially different properties of 3D and 2D systems, a fundamental question that arises when exfoliating magnetic layered compounds concerns the critical thickness below which a material can be considered as 2D. Indeed, the inter-layer exchange coupling $J_L$, although typically very small, leads normally to a 3D critical behaviour of several quantities in sufficiently thick samples. One can then take the dimensional crossover in the critical exponents from the values expected in 3D to those in 2D (see Table 1) to mark the transition from a 3D behaviour to a truly 2D character[17,34]. This is for instance what was done[55] for $Fe_3GeTe_2$. Another fundamental quantity to monitor is $T_c$, which we expect to evolve from its bulk value $T_c^{3D}$ to its 2D value $T_c^{2D}$ as the number of layers is reduced. Theory predicts[56,57] that the critical temperature for a *N*-layer structure, $T_c(N)$, should approach $T_c^{3D}$ for large *N* as

$$T_c(N)/T_c^{3D} = 1 - (C/N)^\lambda \qquad (2)$$

where *C* is a non-universal constant and $\lambda = 1/\nu$ is related to the critical exponent ν of the bulk 3D system (and thus depends on the spin dimensionality; see Table 1). For small *N*, instead, the critical temperature should increase linearly with the number of layers[58]. This prediction describes very well the behaviour of $Fe_3GeTe_2$ where the critical temperature has been measured over a wide range of sample thicknesses[59,60].

Despite having rather general validity, all considerations here above do not exclude the possibility that particular compounds may exhibit a different behaviour due to a variety of specific mechanisms. For example, the situation can be easily imagined in which the evolution of the band structure in a metallic layered material leads to a change in the density of states (DoS) at the Fermi level, which coincidentally exhibits a large peak when the material thickness is reduced to the atomic level. Such a peak in the DoS may enhance the sensitivity to interaction effects, and lead to magnetic order in agreement with the Stoner criterion[61]. Indeed, there are cases in which a ferromagnetic state has been reported in monolayers but not in the bulk, as first experiments on $VSe_2$ appear to indicate[50]. It seems a realistic possibility that the occurrence of magnetism in this case is a consequence of the specific monolayer band structure having an enhanced DOS at the Fermi level[19]. It will be nonetheless important to see whether this very interesting result will be reproduced experimentally, and hopefully extended to more 2D ferromagnetic metals.

Currently a good dozen 2D magnetic crystals have been studied, either obtained by exfoliating 3D parent layered magnets (see Box 1) or by growing them directly on a substrate[50-52]. Due to the limited space available, it is impossible to review all of them here, and we will focus on the two most studied systems - $CrI_3$ and $Fe_3GeTe_2$. Since $CrI_3$ is a semiconductor and $Fe_3GeTe_2$ is a metal – such a selection of materials allows us to overview the whole breadth of the techniques used. We will also focus on the most common features of the magnetic materials, so that much of the information presented can be applied to other materials as well.

**Box 1: Examples of magnetic configurations of vdW layered materials**

**The strong geometrical anisotropy of vdW layered crystals –which leads to a significant difference in magnitude between intra-layer ($J$) and inter-layer ($J_L$) exchange coupling– also manifests itself in different kinds of inter-layer spin alignment found in different material systems. Here we illustrate different examples, along with experimentally-reported materials realisations. Semiconducting and metallic compounds are distinguished with different colours; the scheme does not illustrate the magnetic anisotropy of the different materials (for instance in the bottom right quadrant, $CrI_3$ and $CrCl_3$ are both layered antiferromagnets with all spins ferromagnetically order in each plane, but whereas in $CrI_3$ the layer magnetization points perpendicular to the plane, in $CrCl_3$ it lies in the plane)**

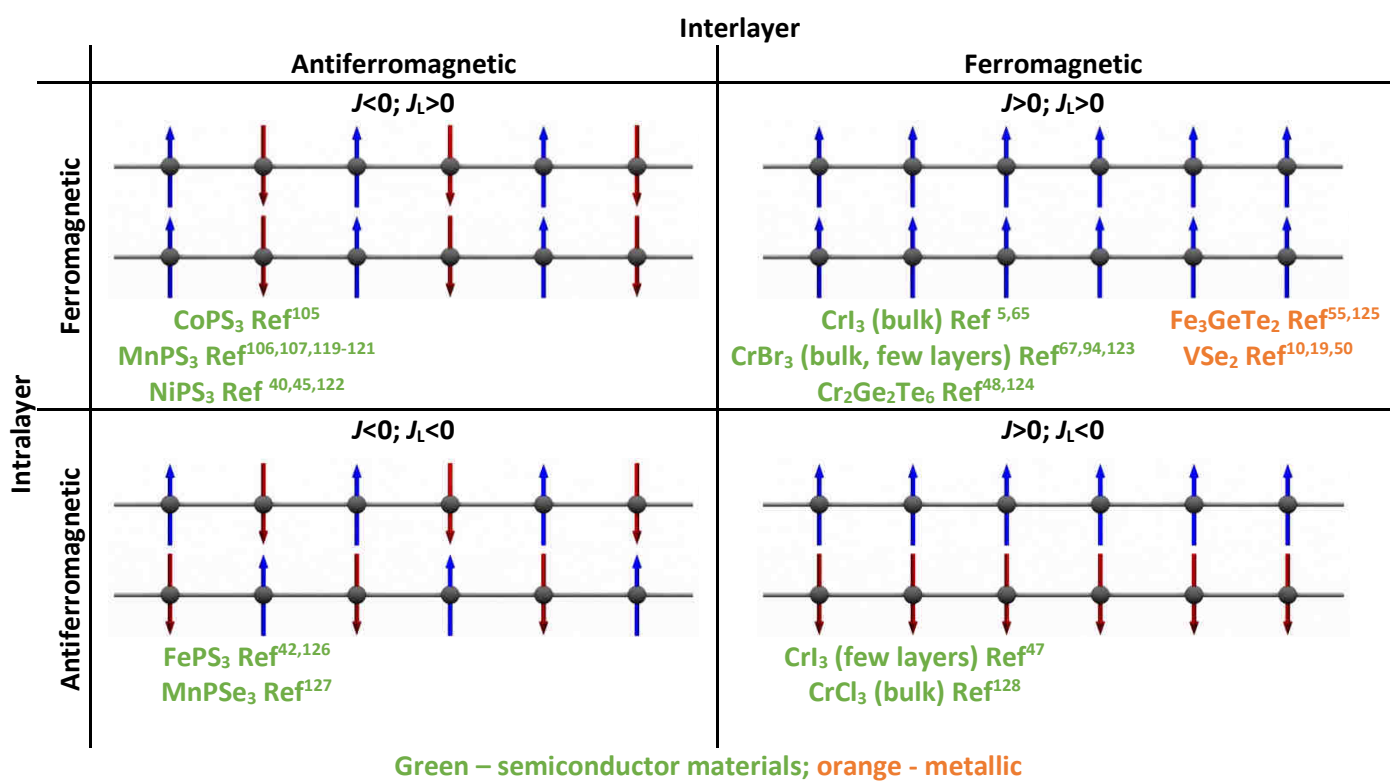

# CrI$_3$

Semiconducting layered van der Waals materials that undergo low-temperature magnetic transitions often exhibit different forms of antiferromagnetic order. Nevertheless, transitions to ferromagnetic states have also been reported in many cases and, quite naturally, early studies of 2D magnetism on exfoliated crystals have relied on compounds with relatively large bulk critical temperature, such as CrI$_3$[5,23,47,62-64].

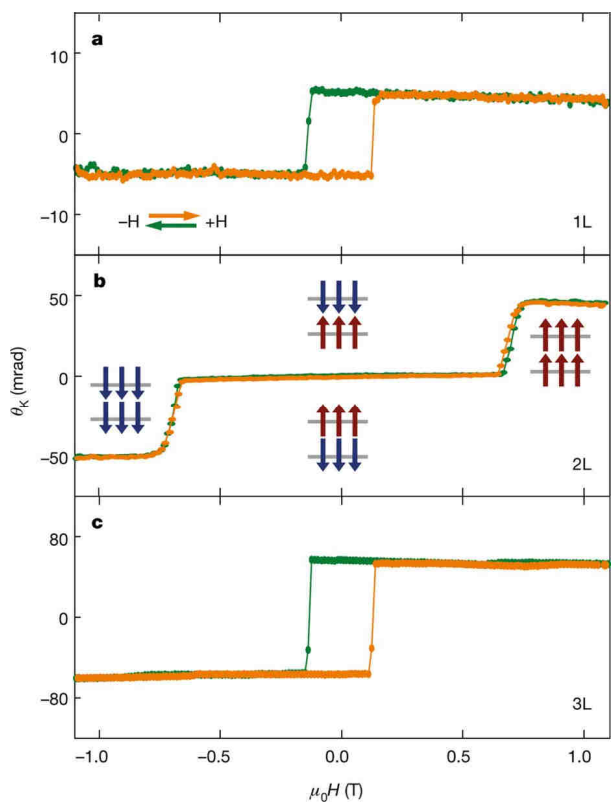

**Fig. 2. Magneto-optical Kerr effect for thin films of CrI$_3$.** The evolution of magnetization with the application of an external magnetic field in out-of-plane configuration may be traced by measuring the Kerr rotation angle. Data for mono-, bi- and trilayer films of CrI$_3$, illustrated in panels **a**, **b** and **c**, unravel the thickness-dependent behaviour of magnetic order in such structures. Samples with an odd number of layers display a hysteresis loop with a coercive field ~0.13 T, indicative of a ferromagnetic state. The bilayer is characterised by ferromagnetic order within the individual layers, with antiferromagnetic coupling between the layers, giving rise at low field to a vanishing Kerr rotation angle. At a field ~±0.65 T a spin-flip process occurs in one of the layers, so that the magnetization of both layers points in the same direction. Adapted from[47].

Experimental investigations of bulk CrI$_3$ and other trihalides date back to sixties[65-71], but only recently McGuire and collaborators have reported a well-documented study of the temperature-dependent structural properties and magnetic response of this material[5]. Magnetization and magnetic susceptibility measurements show that bulk CrI$_3$ is a strongly anisotropic ferromagnet below the Curie temperature ($T_C$=61K), with its easy axis pointing perpendicular to the layers, and a saturation magnetization consistent with a $S$=3/2 state of the Cr atoms. The behaviour observed is typical of a very soft ferromagnet, with the formation of magnetic domains causing the remnant magnetization to vanish and the absence of magnetic hysteresis. Clear evidence for a second magnetic phase

Box 2: The Kerr effect and its application to atomically-thin materials

The magneto-optical Kerr effect (MOKE)[129] is an example of circular birefringence –i.e. a different response in reflection or transmission of a dielectric medium depending on the polarisation state of the incoming photon, which is induced by the presence of magnetic order in the reflecting material. In the case of a specimen exhibiting magnetization along the vertical $z$-axis (polar MOKE), an additional phase difference will arise between $\sigma^+$ and $\sigma^-$ circularly polarised photons travelling along the z-axis upon reflection from the surface. For linearly polarised light, this translates into a rotation of the polarisation direction upon reflection by an angle $\theta_K$, which is known as the Kerr rotation (panel a). In the simple case of a bulk specimen, one can consider only single reflection processes, and the Kerr rotation angle is directly proportional to the magnetization of the sample ($\theta_K \propto M$). However, when adopting MOKE to assess the magnetisation of atomically thin films of layered materials, several difficulties arise. First, the ultimate thinness of 2D magnetic materials and their incorporation into van der Waals heterostructures results in a more complex dependence $\theta_K(M)$, which, to a first approximation, can be interpreted in terms of multiple reflections[47]. Moreover, such macroscopic approach does not consider the character of the electronic structure of the magnetic material (e. g. existence of excitons), which may alter the state of the reflected photons via additional thickness-dependent contributions to the dielectric functions, leading to, e. g., a highly dispersive nature of MOKE signal (see panels b and c for Kerr rotation measurements of the same $CrI_3$ monolayer for 633 nm and 780 nm lasers). Finally, and even more importantly, the description that links the Kerr rotation to the magnetization of the sample is based on Maxwell's equations for macroscopic dielectric media, which is not necessarily always correct for atomically-thin materials. Nonetheless, at present MOKE (together with other polarisation-based optical techniques suitable for probing magnetisation, such as magnetic circular dichroism arising from unequal absorption coefficients for photons characterised by opposite circular polarisation) has been very effective in revealing the magnetic properties of 2D materials.

Panels (b) and (c) are adapted from[40,47].

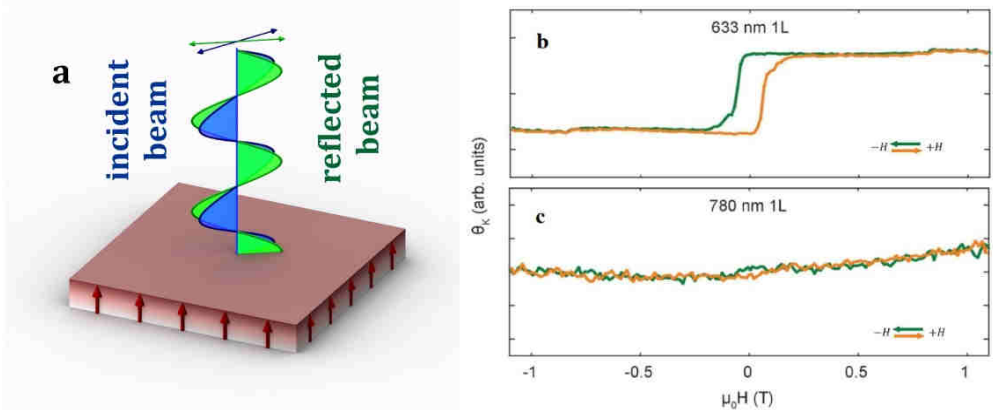

transition at $T \sim 50K$ is found in low-field magnetization measurements with $B$ applied parallel to the layers, but neither its nature nor the details of the ensuing magnetic state are currently understood[72]. This is worth pointing out, because this second transition appears to be related to the unexpected behaviour observed in atomically thin crystals (see below).

The first low-temperature magneto-optical Kerr effect experiments (see Box 2) demonstrating the persistence of magnetism in atomically-thin $CrI_3$ crystals were reported in 2017[47]. The measurements show an unexpected systematics in the magnetic response of $CrI_3$ mono-, bi-, and trilayers, Fig. 2. In mono- and trilayers, a hysteretic Kerr rotation switching at low $B$ (~0.1 T) and saturating at higher $B$ is observed upon sweeping the magnetic field applied perpendicularly to the CrI$_3$ layers. In bilayers, however, the Kerr angle vanishes at low field, and only when sweeping the field over a larger range (B ~ 0.65 T) a sharp, non-hysteretic jump to a state with finite Kerr rotation occurs.

Conventional theory describing magneto-optics of macroscopic media enables a coherent interpretation of these observations. In the measurement configuration, the magnetic field dependence of the Kerr angle indicates that in mono- and tri-layers a finite magnetization perpendicular to the layers is present at B=0, switching hysteretically under an applied field. Mono- and tri-layers are therefore strongly anisotropic ferromagnets with an easy axis perpendicular to the CrI$_3$ layers, as expected from the bulk. The same logic, however, leads to the conclusion that the magnetization of CrI$_3$ bilayers vanishes at B=0, and only appears in the presence of a sufficiently large applied field. This behaviour suggests that the constituent monolayers have equal and opposite magnetization at B=0, so that the total magnetization vanishes. Only when a sufficiently large external field is applied, the magnetic moments in the two layers reorient and point in the same direction, resulting in a finite total magnetization.

The outlined scenario implies that the microscopic interlayer exchange coupling in CrI$_3$ is antiferromagnetic, since only then the low-field bilayer magnetization vanishes. Mono- and trilayers (and more in general all multilayers formed by an odd number of layers) are ferromagnetic because the magnetization on one of the layers remains uncompensated. Such a possibility is internally consistent and agrees with the data, but it is unclear how it can be reconciled with the ferromagnetic state of bulk crystals that seemingly requires the interlayer exchange coupling to be ferromagnetic. Indeed, *ab-initio* calculations based on the actual low-temperature bulk CrI$_3$ structure predict the interlayer exchange coupling to be ferromagnetic[5,62]. It has nevertheless been noticed that the same calculations performed using the high-temperature crystal structure (i.e., the structure stable above a structural phase transition[5] occurring at T ~ 200K) give an antiferromagnetic exchange interlayer coupling[73-77]. This finding has led to the suggestion that the structure of thin CrI$_3$ multilayers may correspond to the high-temperature bulk CrI$_3$ structure and not to the low temperature one. So far, no experimental information is available to support this idea, although clear evidence for a change in magnetic state associated with a structural transition has been recently reported[78] in few-layer CrI$_3$. It remains therefore an open question how (and on which length scale) the crossover from antiferromagnetic to ferromagnetic interlayer coupling occurs as the thickness CrI$_3$ is increased from monolayer to bulk.

## Fe$_3$GeTe$_2$

Another popular 2D ferromagnet is Fe$_3$GeTe$_2$. Unlike CrI$_3$, Fe$_3$GeTe$_2$ is a metal with carrier concentration that has been reported to vary from $1.2\times10^{19}$ cm$^{-3}$ in samples grown by molecular beam epitaxy[51] (with a carrier mobility of ~50 cm$^2$/V·s) to ~$10^{21}$ cm$^{-3}$ (approximately $8\times10^{13}$ cm$^{-2}$ per layer) in crystals grown using the chemical vapour transport method[79]. Alongside Kerr measurements, the anomalous Hall effect has been used to study magnetism[51]. Micromechanical cleavage of this material is difficult using conventional methods and, in order to increase the chances to obtain monolayer flakes, exfoliation was performed on freshly-evaporated gold[55] or on Al$_2$O$_3$[59]. Similarly to CrI$_3$, the strong spin-orbit coupling results in a magneto-crystalline anisotropy along the c-axis (perpendicular to the planes), with a ferromagnetic arrangement of spins within each layer. Although most reports suggest ferromagnetic arrangements between the layers[59], magnetic force microscopy indicates more complex arrangements with an antiferromagnetic phase dominating or coexisting with the ferromagnetic one[80].

When studying the magnetization as a function of thickness, two transitions have been observed. One transition can be seen through the shape of the hysteresis loops. For thin samples (the definition of "thin" depends on the experiment: it corresponds to less than 100 nm in Ref[60] and less than 15 nm in Ref[55]) no domain structure is reported, leading to sharp and perfectly rectangular hysteresis loops, whereas for thicker samples labyrinthine domains appear above a critical temperature and result in a more complex structure of the hysteresis loop.

Another transition can be seen in very thin samples through the deviation of the dependence of $T_c$ on the number of layers from the expected scaling behaviour in equation (1). For samples above ~5 layers, the thickness dependence follows equation (1), with the constant C corresponding to approximately 2-3 layers[59,60] and λ of the order of 1.7 – close to what is expected for 3D Heisenberg ferromagnetism[59,60]. Deviations from this behaviour occur for thinner samples, which has been taken as an indication for a transition from 3D to 2D behaviour[59]. More evidence for such 3D to 2D transition comes from the careful study of the critical behaviour of the magnetization as a function of temperature. For samples above 5nm in thickness the magnetization dependence on temperature is well described by the power law $M(T)=M(0)(1-T/T_c)^\beta$ with β in the range of 0.25-0.27. For thinner samples the critical exponent gradually decreases, reaching 0.14 for a monolayer[55], close to the value expected for the 2D Ising model (see Table 1).

Magnetic circular dichroism[55] and anomalous Hall effect[59] measurements have shown that magnetism survives down to the monolayer limit, with $T_c$ dropping down to 130K and 20K, respectively. At this moment of time it is difficult to establish the origin of this discrepancy, because monolayers of this material are difficult to produce, and thus rather unconventional fabrication methods are being used[59] which might lead to changes of the properties of the resulting devices (either due to introduction of defects or due to the unintentional doping). For instance, slight changes in the iron content are known to affect significantly the critical temperature of the bulk[81]. Also, the monolayers studied in Ref. [55] were in direct contact with a gold substrate, whereas those studied in Ref. [59] were not, so that substantial charge transfer could have taken place in the first case and not in the second. As we know that $T_c$ is very sensitive to the electron density - this difference may have had a large influence.

## Gating of atomically thin magnetic crystals

Electrostatic control of magnetism in gated structures is of great interest for fundamental reasons and for its potential for future device applications[82]. In thin films of conducting materials, gate-controlled magnetism has been demonstrated over the last two decades, first in magnetically doped semiconductors and subsequently in conventional ferromagnetic metals[83-86]. Gating typically causes changes in the critical temperature and in the coercive field, which are driven by the accumulated charge carriers[59]. The possibility to influence electrostatically the magnetic properties of so-called multiferroics, i.e. systems in which the magnetization and the electrical polarization are intrinsically coupled, is also well-established[87,88]. In this case, changes in magnetization are driven by the applied electric field and not by the carrier density, a phenomenon normally referred to as magnetoelectric effect.

Establishing the possibility to gate-tune the magnetic properties of atomically thin crystals has been a prime goal of experiments on virtually all magnetic 2D materials investigated so far. For $CrI_3$ and $Fe_3GeTe_2$, different experiments have been performed to show the influence of gating on magnetism. For semiconducting $CrI_3$, work has relied on MOKE measurements in singly and double gated devices, whereas for metallic $Fe_3GeTe_2$ the anomalous Hall resistance has been measured.

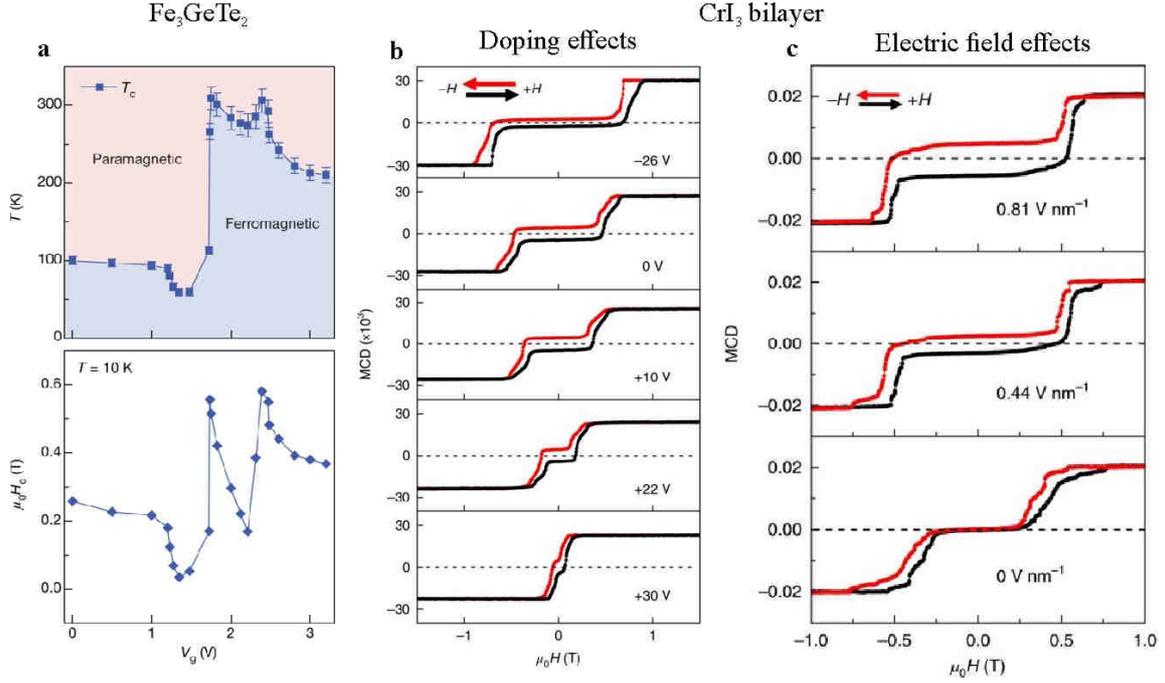

**Fig. 3. Doping and gate control of layered magnets. a, b** Measurements of the anomalous Hall effect in trilayer $Fe_3GeTe_2$ demonstrate the possibility to tune the Curie temperature with ionic gating. **a**, The resulting phase diagram shows that the transition from the ferromagnetic to the paramagnetic state strongly depends on the gate voltage, and that the Curie temperature may be increased as high as up to room temperature. **b**, The coercive field characterising $R_{xy}$ hysteresis loops at low temperatures also displays a strong dependence on the ionic gate voltage, which correlates with the dependence of $T_c$. **c, d** Double gated structures allow for separate investigation of the doping-related and electric-field-related changes in the magnetic state of bilayer $CrI_3$. **c**, The magnetization evolution, traced by measurements of magnetic circular dichroism (MCD), is presented for different doping levels. Electron/hole doping lowers/increases the value of magnetic field required to break the interlayer antiferromagnetic coupling. **d**, The three panels present the impact of an electric field on the magnetic state of $CrI_3$ bilayer. The major effect arising from the presence of the electric field is the opening of a hysteresis loop centred around zero magnetic field, indicative of a non-zero magnetization in the antiferromagnetic phase. Panels **a** and **b** are reproduced from[59], panel **c** - from[90], and panel **d** - from[118].

Work on $Fe_3GeTe_2$ done at Fudan University illustrates the possibility to control the Curie temperature of atomically thin layers over an unexpectedly large range[59]. This was shown by measuring the anomalous Hall resistance of trilayer $Fe_3GeTe_2$ devices gated with a lithium-based electrolyte (note, however, that there is a possibility that apart of pure electrostatics, there are also effects due to intercalation). At zero gate voltage, the trilayer $T_c$ determined from the disappearance of the hysteresis in the $R_{xy}(B)$ curve was found to be $T_c \sim 100$ K, much lower than the bulk value. Upon applying a positive gate voltage, however, $T_c$ was found to vary in a non-monotonic way, reaching values in excess of 300 K. Interestingly, the evolution of the coercive field $H_c$ extracted from fixed-temperature gate-dependent measurements parallels that of the critical temperature, with $H_c$ and $T_c$ exhibiting maxima and minima at the same gate voltage values, Fig. 3a,b. The phenomenon was attributed to the shift in the Fermi energy $E_F$ of $Fe_3GeTe_2$ –with a concomitant change in DoS at $E_F$- and used to interpret the data in terms of Stoner's criterion for itinerant ferromagnetism[61]. Whereas the experimental observations are certainly tantalizing and the proposed interpretation plausible, a more comprehensive characterization of the experimental system as a function of gate voltage will be needed to establish definite conclusions. For instance, it will be essential to measure the gate-induced charge density, to understand whether or not lithium intercalates, and to consider other effects that lithium may have besides causing charge transfer (e.g., on the $Fe_3GeTe_2$ crystal structure).

Gate-dependent magnetism in $CrI_3$ has been investigated by means of magneto-optics (i.e., gate dependent Kerr effect measurements) in structures with conventional solid-state gates. The most interesting results have been obtained on double-gated bilayer $CrI_3$ devices, i.e., structures in which an hBN-encapsulated $CrI_3$ bilayer connected to graphene electrodes is sandwiched between two conducting materials acting as gates. Different information is obtained when biasing the two gates with the same or opposite polarity. If the voltage polarity is opposite, the $CrI_3$ bilayer can be kept at zero total accumulated charge (the charge accumulated by one gate is removed by the other) while imposing a perpendicular electric field of varying strength, proportional to the difference of applied voltages. Alternatively, if voltages of the same polarity are applied, the perpendicular electric field can be made to vanish, so that the only effect of gating is to accumulate charge. Double-gated devices, therefore, allow the influence of an applied perpendicular electric field and of accumulated charge carriers to be discriminated experimentally.

The application of a perpendicular electric field and the accumulation of charge have strikingly different effects, which can again be monitored by Kerr effect measurements. Doping of $CrI_3$ crystals (through biasing the two gates with voltages of the same polarity) causes a progressive reduction in the value of magnetic field (Fig. 3c) that is needed to switch the $CrI_3$ from the antiferromagnetic state stable at low field, into a ferromagnetic state stable at high field (i.e., the state in which the magnetization of the two $CrI_3$ monolayers have the same orientation). The same phenomenon has been reported by two independent teams of researchers, albeit the magnitude of the observed reduction in switching field differ significantly in the two cases: one team reported[89] the switching field to decrease by about 30%, whereas the other found the switching field can be made to vanish altogether[90]. This is a quite remarkable finding, as it implies that charge accumulation actually turns bilayer $CrI_3$ from an antiferromagnetic into a ferromagnetic state at $B=0$.

The application of a perpendicular electric field results in a distinctly different modification of the magnetic properties. As mentioned earlier, in an extended interval of applied magnetic field around $B=0$ (up to values $B\sim \pm 0.5$-$0.7$ T, depending on the device) the total magnetization of bilayer $CrI_3$ vanishes, due to the antiparallel orientation of the spin in the adjacent layers, and thus, no Kerr rotation is observed. When applying a perpendicular electric field ($E_z$), however, a finite and hysteretic Kerr rotation appears in this same low magnetic field range, with the Kerr angle that increases linearly with $E_z$, Fig. 3d. If interpreted in terms of conventional magneto-optics theory, this observation implies that the applied electric field induces a finite magnetization $M_z \propto E_z$, and thus the presence of a strong magneto-electric effect in bilayer $CrI_3$.

The overall consistency of the qualitative aspects of the results reported in different experiments gives confidence in the robustness of the observed phenomena. Much remains to be understood, however, about the precise conditions of the experiments and the underlying microscopic physics. For instance, it is currently unclear how large is the accumulated charge density when the two gates are biased with the same voltage polarity and – since the Fermi level is inside the band-gap of $CrI_3$– which electronic states are occupied by the accumulated charge. Some caution is also desirable when interpreting the effect of the applied electric field, because theory predicts[91] that when the combination of inversion and time-reversal symmetry is broken (which is the case in $CrI_3$ bilayers in the presence of a perpendicular electric field) Kerr rotation can occur in an antiferromagnet with vanishing magnetization. Hence, the observation of finite Kerr rotation does not always imply that a finite magnetization is present and a direct measurement of the electric field induced magnetization would be desirable.

## Heterostructures

The use of magnetic 2D materials as part of van der Waals heterostructures pursues two targets. One is expanding the functionality of the heterostructures available for experiments and for future applications. The other is to use phenomena originating from proximity to other 2D crystals as a tool to study the magnetic properties of the magnetic layers themselves.

### Magnetic tunnel junctions

Due to the absence of broken covalent bonds, the surfaces of layered van der Waals bonded materials normally possess high-quality electronic properties. This enables one to realise magnetic tunnel junctions by sandwiching thin films of insulating or semiconductor materials (like hBN or TMDC) between two layers of metallic ferromagnetic crystals, as has been done recently with $Fe_3GeTe_2$[92]. A pronounced spin-valve effect has been observed in $Fe_3GeTe_2$/hBN/$Fe_3GeTe_2$ junctions due to different coercive field of the top and bottom electrodes (because of different demagnetisation factors for the two electrodes with different shape), Fig. 4a. The observed 160% magnetoresistance corresponds to a 66% spin polarization at the Fermi energy, i.e., to the presence of 83% and 17 % of majority and minority spins, respectively. These experiments also confirmed that the magnetic properties of the surface of $Fe_3GeTe_2$ crystals are very similar to those of the bulk.

A less conventional type of magnetic tunnel junctions can be realized by using exfoliated insulating magnetic 2D materials as tunnel barriers, in combination with non-magnetic electrodes (typically graphene multilayers). Two basic mechanisms leading to magnetoresistance in such structures have been identified. One is magnon-assisted tunnelling: at sufficiently high-bias (typically 10 mV or higher) additional inelastic tunnelling channels open up, in which tunnelling electrons emit another quasiparticle, thus increasing the differential conduction, Fig. 4b. In the case of non-magnetic barriers (and for magnetic barriers above $T_c$) such effects are dominated by phonon emission[93]. However, for $T<T_c$ electrons can also emit magnons, identifiable by the shift of their energy upon applying magnetic field[94,95].

Much larger effects can originate from switching of the magnetic state in the barrier, as observed in $CrI_3$. Due to the antiferromagnetic coupling of the magnetization in the adjacent layers, $CrI_3$ tunnel barriers exhibit negative magnetoresistance in excess of 10000%[72,73,96], occurring in a series of sharp jumps, Fig. 4c-e. The phenomenon can be understood by considering that in the antiferromagnetic state each next layer acts as a tunnel barrier for electrons with the opposite spin. In the antiferromagnetic state, therefore, the tunnelling resistance is high for both spins. In the ferromagnetic state, when the magnetization in all layers is aligned, the barrier is high for the minority spins and much lower for the majority spins, which reduces the overall resistance of the structure. Since the different layers in the multilayer $CrI_3$ switch their magnetization at different magnetic field, multiple jumps occur, each corresponding –in the simplest scenario– to one layer reverting its magnetisation, as it can be inferred by comparing the tunnelling magnetoresistance to the magnetic field evolution of the MOKE measured on the same devices.

Furthermore, thin multilayers $CrI_3$ with the same overall magnetisation, but different magnetisation sequence of the individual monolayers (e.g., ↑↓↑↑ and ↑↑↓↑ in a tetralayer) can exhibit different magnetoresistance[97,98]. In practice, even for thick $CrI_3$, usually two magnetoresistance jumps at approximately 0,85 and 1.9 T are found to dominate the switching virtually in all devices, up to crystals exceeding 10 nm (in which tunnelling occurs in the Fowler-Nordheim regime under the application of high bias[73]). The reason why these two jumps in magnetoresistance are so much larger and more robust than all the others remains to be understood in detail. Still, it is noteworthy that the values of the magnetic field at which these most prominent jumps occur correspond precisely the values in which the exchange field induced by a 20-nm $CrI_3$ crystal into a $WSe_2$ monolayer is also seen to switch (see below).

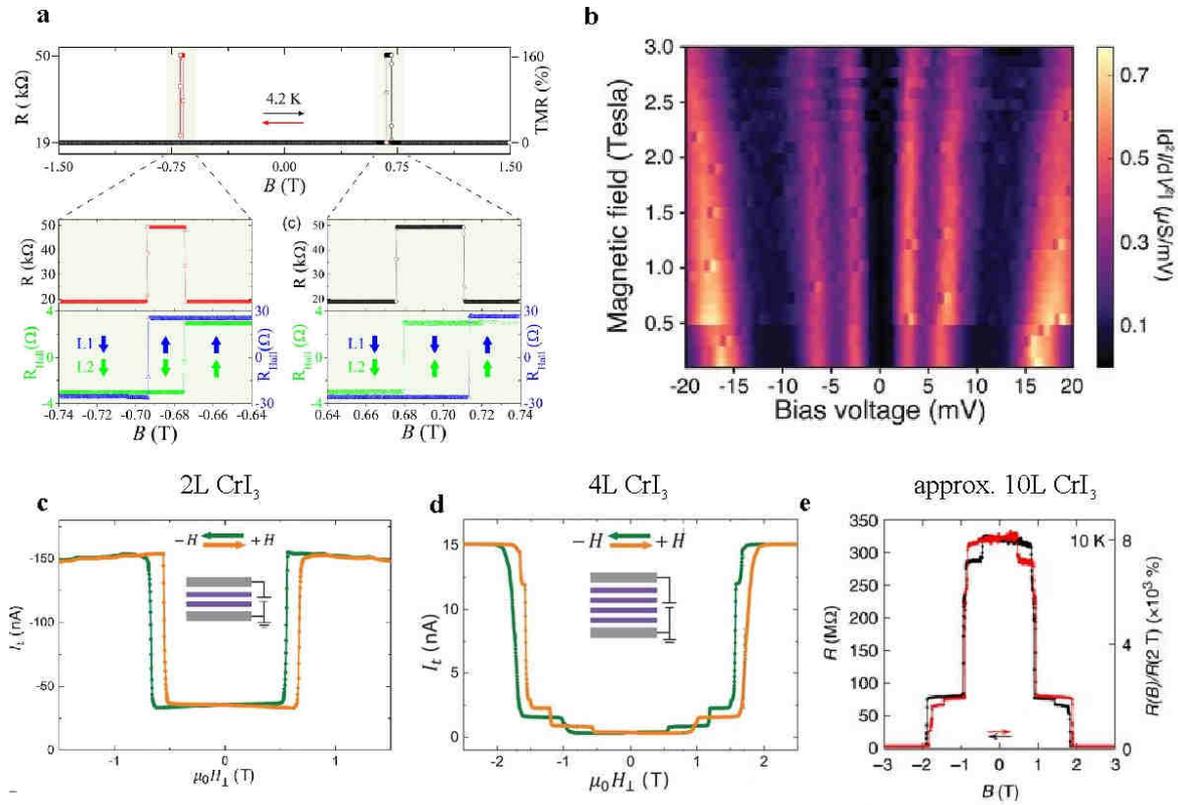

**Fig. 4. Heterostructures with layered magnetic materials a**, Ferromagnetic layered metals may be employed as magnetic contacts, enabling fabrication of devices acting as spin filters. The data presented show the magnetic field evolution of resistance for a $Fe_3GeTe_2$/BN/$Fe_3GeTe_2$ junction which displays a spin-valve behaviour with magnetoresistance of up to 160% arising from a difference in conductance when the magnetization of the two $Fe_3GeTe_2$ electrodes is parallel or antiparallel. **b**, Differential tunnelling conductance for Graphite/$CrI_3$/Graphite junctions measured as a function of bias voltage and magnetic field applied along the magnetization easy axis of $CrI_3$ reveals low-energy resonances exhibiting a linear dependence on the magnetic field, consistently with the expected evolution of magnon states in $CrI_3$. The emergence of such inelastic tunnelling process provides a way to explore low-energy collective states in layered magnetic materials. **c-e**, Vertical transport experiments in tunnel junctions based on thin layers of $CrI_3$. The impact of magnetic field on the tunnelling current of 2-, 4- and ~10-layers $CrI_3$ devices is presented in panels **c**, **d** and **e**, respectively. The emerging current/resistivity jumps at certain values of magnetic field correspond to changes in the magnetic state of $CrI_3$. For the bilayer device, the sharp resistivity step arises from breaking the interlayer exchange coupling. The 4-layers device exhibits a more complex evolution of the tunnelling current, which may be attributed to layer-by-layer spin-flip processes, leading to the appearance of multiple steps. In the 10-layers sample, two major jumps of resistance arise. Panel **a** is reproduced from[65,92], **b** - from[95], **c** and **d** - from[96] and **e** - from[45,73].

Current works are exploring different aspects of $CrI_3$ tunnel barriers to realize transistors with magnetic functionality. These include, for example, the dependence of the tunnelling magnetoresistance on the position of the Fermi level in the graphene contacts and the effect of the accumulated charge on the magnetic state of $CrI_3$[96-98]. Investigations are also attempting to understand how the magnetic behaviour crosses over from that of exfoliated crystals to that of the bulk, a highly non-trivial question since the behaviour of exfoliated CrI$_3$ crystals that are 10-20 nm – seemingly thick enough to be considered as bulk– is similar to that of atomically-thin crystals and not to that of truly macroscopic samples.

## Heterostructures with semiconducting TMDCs

Proximity between magnetic 2D crystals with other layered materials may serve as a very sensitive tool to study the details of the magnetic structure. The first investigations demonstrating pronounced magnetic phenomena at van der Waals interfaces were focused on CrI$_3$/WSe$_2$ structures[63]. In such devices, the low-temperature photoluminescence of WSe$_2$ reveal zero-field Zeeman-type splitting of excitonic resonances. The effect is due to an interfacial exchange energy equivalent to the presence of an effective 13T magnetic field. The fact that the effect disappears above the bulk Curie temperature of CrI$_3$ confirms the origin of the phenomena. These first experiments also probed the evolution of the exchange field with applied perpendicular magnetic field, which was found to be more complex than expected for a conventional magnet. One of the most pronounced features in the photoluminescence measurements is a sharp jump at *B* ~ 0.85 T, indicative of a field-induced change in the magnetic state of the system. The position of the jump corresponds well to those observed in the tunnelling experiments, calling for further investigation of the phenomena.

## Outlook

The field of 2D magnetic materials has just started, but the experimental results already reported allow us to catch the first glimpse of a vast unexplored domain. Based on the work done until now, we can identify different questions that require urgent attention.

Broadening the experimental investigations to a much wider spectrum of compounds is urgently needed (many candidate materials are available, with a clear predominance of layered semiconducting systems). This would allow us to better structure our knowledge of the behaviour of the layered ferromagnetic materials, and to understand microscopically how this behaviour is determined by the strength on inter- and intra-layer exchange interactions, uniaxial anisotropy, *etc*.

It will be instructive, for instance, to find more semiconductors with an out-of-plane layer-antiferromagnetic spin configuration, and check if giant tunnelling magnetoresistance always occurs as it does in CrI$_3$. It is not obvious that it should be so, because the coupling between the spin configuration and the electronic states mediating tunnelling need not be the same in different compounds. Indeed, the description of magnetic 2D materials requires the introduction of a variety of microscopic physical parameters, such as in-plane and out-of-plane magnitude of exchange interactions, different spin-orbit coupling terms and their strength, etc. How these parameters determine the magnetic configuration of the system and influence the single-electron states responsible for the opto-electronic response of different 2D magnetic materials is far from being well-understood. Comparing systematically the behaviour of materials with different microscopic parameters is the only way to the progress. These considerations should be extended to van der Waals interfaces and heterostructures of different 2D magnetic materials, whose investigations enable atomically thin crystals to be probed in unique ways and have a great potential for unexpected discoveries.

Enlarging the family of magnetic monolayers entails also the possibility to provide experimental evidence for long-standing theoretical predictions that still lack validation in truly 2D systems. This includes the realization of exotic states of matter like intrinsic Chern insulators[99] and quantum spin liquids[100], which have currently been observed only in non-stoichiometric magnetically-doped materials[101] or in 3D layered crystals not yet exfoliated into monolayers[102], respectively. Moreover, strong Dzyaloshinskii-Moriya interactions could stabilize topological spin configurations known as skyrmions that, in addition to their fundamental interest, could provide a new paradigm for high-density, low-power data storage[103].

Future experiments will additionally require significant technical development and new ideas to probe magnetism quantitatively on very small length scales and in the conditions of very large demagnetisation fields. This is certainly needed, for instance, for the investigation of antiferromagnetic 2D materials, with adjacent spins in a same layer pointing in alternating directions. A number of these compounds have been studied as bulk cystals[104,105] –$MnPS_3$, $FePS_3$, $CoPS_3$ – but only some of them have been exfoliated down to monolayer[106-108]. In the bulk, antiferromagnetic order is probed by neutron diffraction experiments, which – due to the weak interaction of neutrons with matter – requires the use of very large bulk crystals. For antiferromagnetic atomically thin crystals new strategies will have to be followed. The use of Raman spectroscopy[42] to detect antiferromagnetic ordering in thin (down to monolayer) crystals of $FePS_3$ illustrates an idea that has been pursued recently. We anticipate that in the future different types of scanning probes will play an important role. Examples are spin-polarized scanning tunnelling microscopes, which possesses the required atomic resolution, and scanning magnetometers based on NV-center or nano-SQUIDs that –albeit not sensitive to atomic scale details– provide quantitative measurement of the magnetization with spatial resolution reaching few tens of nanometers.

The considerations made here show that there are many different important directions for future work. It is clear that the results reported during the last one or two years represent the starting point of a new field of research in which major future developments should be expected. At this stage, the key questions that are being addressed are of fundamental nature, but as soon as 2D magnetic materials will be reliably synthesized with sufficiently high critical temperatures, the potential for technological impact is enormous.

# Acknowledgements.

We would like to acknowledge useful discussions with (in alphabetical order): I. Gutiérrez-Lezama, H. Henck, G. Long, N. Ubrig, Z. Wang. This work was supported by the EU Graphene Flagship Program (K.S.N. and A.F.M.), European Research Council Synergy Grant Hetero2D (K.S.N.), the Royal Society, Engineering and Physical Research Council (EPSRC UK, K.S.N), US Army Research Office (W911NF-16-1-0279, K.S.N. and M. K.), Swiss National Science Foundation Ambizione grant program (M.G.) and Division II (A.F.M.).

# Competing interests

The authors declare no competing interests.